\documentclass[modern,trackchanges]{aastex63}
\usepackage{amssymb, amsmath}
\usepackage{color}
\usepackage{morefloats}
\usepackage{hyperref}
\begin{document}

\title{A Survey for C\,{\sc ii} Emission-Line Stars in the Large Magellanic Cloud}

\author{Bruce Margon}
\affiliation{Department of Astronomy \& Astrophysics, University of California, Santa Cruz, 1156 High St., Santa Cruz, CA 95064, USA} \email{margon@ucsc.edu}

\author{Philip Massey}
\affiliation{Lowell Observatory, 1400 W Mars Hill Road, Flagstaff, AZ, 86001, USA}
\affiliation{Department of Astronomy and Planetary Science, Northern Arizona University, Flagstaff, AZ, 86011-6010, USA}
\email{phil.massey@lowell.edu}

\author{Kathryn F. Neugent} 
\affiliation{Lowell Observatory, 1400 W Mars Hill Road, Flagstaff, AZ, 86001, USA}
\affiliation{Department of Astronomy, University of Washington, Seattle, WA, 98195, USA}
\email{kneugent@uw.edu}

\author{Nidia Morrell}
\affiliation{Las Campanas Observatory, Carnegie Observatories, Casilla 601, La Serena, Chile}
\email{nmorrell@carnegiescience.edu}

\begin{abstract}

We present a narrow-band imaging survey of the Large Magellanic Cloud, designed to isolate the C\,{\sc ii} $\lambda\lambda$7231, 7236 emission lines in objects as faint as $m_{\lambda7400}\sim18$. The work is motivated by the recent serendipitous discovery in the LMC of the first confirmed extragalactic [WC11] star, whose spectrum is dominated by C\,{\sc ii} emission, and the realization that the number of such objects is currently largely unconstrained. The survey, which imaged $\sim$50$~$deg$^2$ using on-band and off-band filters, will significantly increase the total census of these rare stars. In addition, each new LMC [WC] star has a known luminosity, a quantity quite uncertain in the Galactic sample. Multiple known C\,{\sc ii} emitters were easily recovered, validating the survey design. We find 38 new C\,{\sc ii} emission candidates; spectroscopy of the complete sample will be needed to ascertain their nature.  In a preliminary spectroscopic reconnaissance, we observed three candidates, finding C\,{\sc ii} emission in each.  One is a new [WC11].  Another shows both the narrow C\,{\sc ii} emission lines characteristic of a [WC11], but also broad emission of C\,{\sc iv}, O\,{\sc v}, and He\,{\sc ii} characteristic of a much hotter [WC4] star; we speculate that this is a binary [WC].  The third object shows weak C\,{\sc ii} emission, but the spectrum is dominated by a dense thicket of strong absorption lines, including numerous O\,{\sc ii} transitions. We conclude it is likely an unusual hot, hydrogen-poor post-AGB star, possibly in transition from [WC] to white dwarf. Even lacking a complete spectroscopic program, we can infer that late [WC] stars do not dominate the central stars of LMC planetary nebulae, and that the detected C\,{\sc ii} emitters are largely of an old population.

\end{abstract}

%\keywords{planetary nebula: individual} 

%\received{2020 April xx}
%\revised{2020 May xx}
%\accepted{2020 xxx}

% UAT "concepts" entries will be prompted during submission process
%{Planetary nebulae nuclei (1250),  WC stars(1793)}	

\section{Introduction}

Although most central stars of planetary nebulae are hydrogen-rich, a modest subset ($<$10-20\%) are not, and instead show intense emission lines of carbon and helium, superficially similar to that of Population I (high mass) Wolf-Rayet stars (see, e.g., \citealt{demarco02,DePew,Todt}).  These lower-mass stars are generically referred to as ``[WR]" or more specifically as ``[WC]" following the suggestion of \citet{vanderHucht81}.  The lowest excitation class of these is designated as [WC11] in the classification scheme of \citet{crowther} and \citet{acker}, and the dozen or so known examples have spectra dominated by
dozens of intense, narrow C\,{\sc ii} emission lines.

Recently \citet{Margon20} (hereafter ``M20’’) reported the serendipitous discovery of a previously unnoted [WC11] star, UVQS~J060819.93-715737.4 (hereafter ``J0608’’), which proves to be a member of the Large Magellanic Cloud (LMC). That paper also discusses the past history of work on these curious objects, as well as their evolutionary status.  The previously anonymous nature of J0608, despite its relative brightness ($V\sim15$) and spectacular emission spectrum, may lead one to ponder how common these ``rare” objects might actually be, especially as at this magnitude, the LMC has been repeatedly and exhaustively surveyed for emission-line stars. If one such object has been missed, perhaps many others also exist. As none of the Galactic examples is close enough for a meaningful {\it Gaia} parallax, each and every LMC [WC11] star is a valuable luminosity calibrator for these exotic objects. M20 have pointed out that the intensity of C\,{\sc ii} lines probably requires a preferential excitation mechanism, not yet understood, so further examples may help to clarify the atomic physics involved. Finally, it is unclear if metallicity differences between the Milky Way and the LMC might impact the occurrence frequency or nature of these stars. 

The very strong, narrow C\,{\sc ii} emission lines are a completely unique feature of this class of star, and therefore are an obvious search criterion. The modest angular size of the LMC at these magnitudes enables a comprehensive wide-field imaging survey optimized for detection of C\,{\sc ii} emission to be conducted even with small telescopes. Here we describe such a survey, together with initial results.

\section{The Survey}
We modeled our survey to discover additional [WC11] stars in the LMC on the recently completed survey for Population I WR stars in the Magellanic Clouds by three of the present authors \citep{NeugentMCIV}. As in that survey, we used interference filters to differentiate candidates from their neighboring stars, employing a combination of photometry and image subtraction.  Given that J0608 is bright, a large aperture telescope was not needed, so we again utilized the wide-field capabilities of the Las Campanas Observatory (LCO) 1-m Swope telescope and its nearly quarter-square-degree field-of-view CCD camera.  In this section we discuss the design and execution of the imaging survey, and the identification of C\,{\sc ii}-emission candidates.

\subsection{Design of the Filter System}

\begin{figure}
\plotone{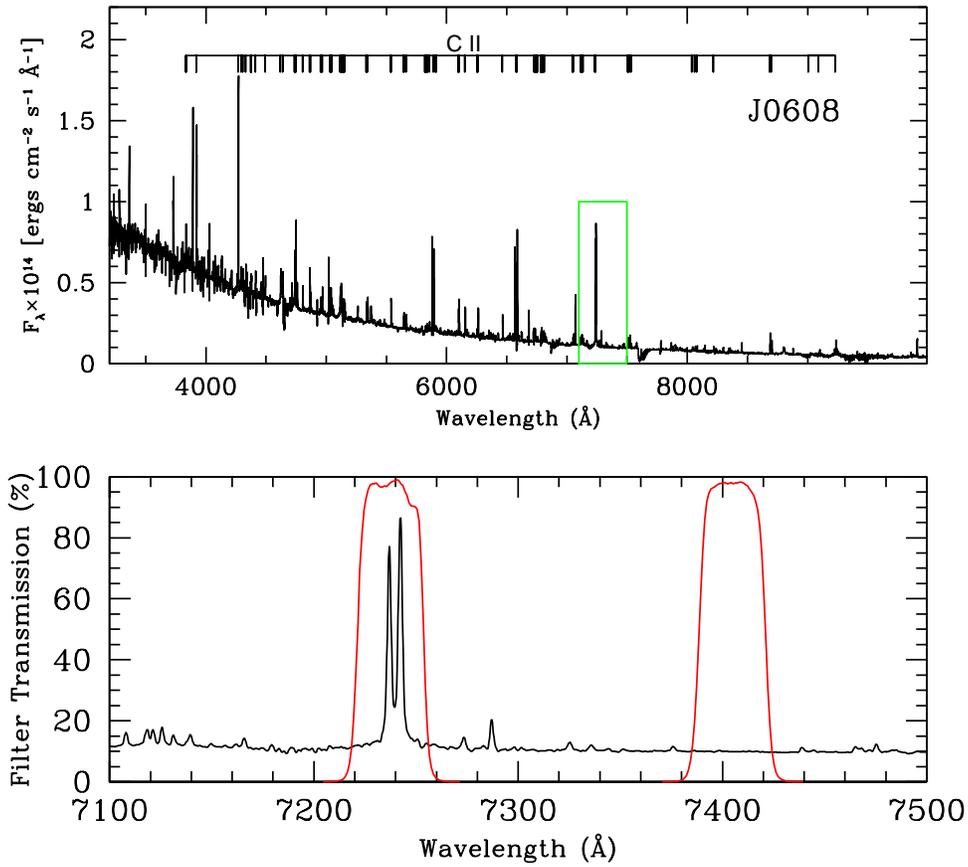}
\caption{\label{fig:filters}  {\it Upper:} The flux-calibrated spectrum of the LMC [WC11] star J0608, from \citet{Margon20}.  The green box isolates the region shown in the lower figure.  {\it Lower:} The laboratory-measured transmission curves for the C\,{\sc ii} on-band filter ($left$) and the continuum off-band filter ($right$) are shown in red, superposed on a section of spectrum of the upper figure. A correction for the wavelength shift due to the effective {\it f}-ratio of the Swope telescope has been included.}
\end{figure}

To identify more potential [WC11] stars we used a system of two interference filters, one centered on C\,{\sc ii} emission, and one on the neighboring continuum.  The goal was to be sensitive to C\,{\sc ii} emission, yet produce few false positives. The M20 fluxed spectrum of J0608 was crucial for this effort.  As shown in Fig.~\ref{fig:filters}, the strongest lines are C\,{\sc ii} $\lambda 4267$ and the C\,{\sc ii} $\lambda\lambda 7231, 7236$ doublet.   We chose to center the interference filter around the latter, in part because the combined strength is slightly greater (see Table~1 of M20).  It is also the strongest line in CPD $-56^\circ$8032, the prototype of the [WC11] class, as shown in Fig.~2 of M20.  Equally importantly, the region redwards of the  $\lambda\lambda 7231, 7236$ doublet is essentially emission line-free, making it suitable for placement of the continuum filter. We chose 7240\,\AA\ as our optimal central wavelength for the on-band filter, allowing for the 262~km s$^{-1}$ systemic radial velocity of the LMC \citep{vandermarel}.

The choice for the center for the continuum filter was less crucial.   As argued elsewhere \citep{AM85, MasseyMCWRI} it is very useful in such detection work to have the continuum filter slightly redwards of the on-band filter rather than bluewards.  This reduces the number of red stars showing up as false positives simply because of their colors. Red stars are much redder than blue stars are blue; i.e., they have a much stronger gradient in flux with wavelength.  Thus, a continuum filter placed $\sim100$\,\AA\, to the red is unlikely to produce a false positive simply because of a star's color,  whereas one placed 100\,\AA\, to the blue may do so.   We therefore chose the continuum bandpass to avoid lines in J0608, with some attention to also avoid any strong OH night-sky emission bands \citep{Osterbrock96}.  (There is some OH emission in the on-band filter that cannot be avoided.)  For the off-band central wavelength we settled upon 7410\,\AA.

Other relevant design specifications were the width of the bandpass, and the general shape. The more narrow the bandpass, the greater the contrast between on-band and off-band, but this requirement must be balanced with the rotational velocity of the LMC, which creates a dispersion in stellar radial velocities.   Measurements of yellow and red supergiants by \citet{Neugent12} show an average radial velocity of 260~km~s$^{-1}$, with a spread of roughly $\pm$80~km~s$^{-1}$.  Using the fluxed J0608 spectrum, we performed simulations of the implied magnitude difference, given different filter shapes (Gaussian vs.\ square-well) over a range of $\pm$100~km~s$^{-1}$. We found that 30--35\,\AA\ wide bandpasses provided uniformly good coverage in this parameter space.  A square-well filter provided far better contrast than a Gaussian, with the extended wings in the latter case picking up additional unwanted continuum flux.   We also note that it is technically difficult (and expensive) to make filters with narrower bandpasses than these that are also uniform over the entire area. 

Using these filter specifications, our expectation was that we would find an on-band minus off-band magnitude difference of $\sim$0.9~mag for J0608.  Based upon the measured flux, and expected throughput of the system, we estimated that even a 300\,s exposure would yield $\sim$22,000 $e^-$ in the on-band vs.\ 10,000 $e^-$ in the off-band, a $75\sigma$ detection above background.
%\footnote{In practice, we assume a very conservative 0.02~mag floor for the errors in quantifying the significance levels of our detections.}.

The 75 $\times$ 75\,mm interference filters were manufactured to our specifications by the Chroma Technology Corporation, using a magnetron thin film sputtering process.  Laboratory measurements of the resulting filters indicated extremely high throughput and a very sharp square-well design, as shown in Fig.~\ref{fig:filters}, but such measurements are obviously no substitute for data from the telescope. Our first exposure in the actual survey was, of course, the star J0608, which motivated this work. We were relieved to measure 22,500 $e^-$ for this star in the on-band and 9,800 $e^-$ in the continuum in 300~s exposures, much as predicted.

\subsection{Observations and Reductions}

The previous Population I WR survey mentioned above covered most of the current star-forming regions of the Large Magellanic Cloud as judged from H$\alpha$ imaging \citep{MasseyMCWRI} and extended out a radius of 3.5$^\circ$.  However, J0608 is located 4.3$^\circ$ from the optical center of the LMC.  This is consistent with the premise that it is part of an older population, and the relatively recent recognition that the full size of the LMC extends far beyond the current star-forming regions \citep{Saha10, nidever}.  We therefore decided to image much farther out than in our previous WR survey, optimally extending our coverage to a radius of $5^{\circ}$ or more.  That would have required 345 fields, allowing for 1$\arcmin$ overlap between adjacent fields. As this was not entirely practical, we contented ourselves with only sampling some of the most distant fields, guided by the number of 2MASS stars present.

The surveyed regions are shown in Fig.~\ref{fig:survey}. In all, 245 fields were observed, as well as ten repeated exposures of a 246th field that was
centered on J0608.  The imaging was performed during two
observing runs, from UT 2019 December 5-9 and 2020 January 8-14. Conditions were quite good on all nights, except for 8 January when the telescope was intermittently closed due to humidity or high winds, and 11 January, when clouds interfered with most observations.  Fields observed through clouds on that night were reobserved later in the observing run. Some high altitude ash from Australian wild fires was evident at sunset during the January nights, but based on repeated observations of J0608 throughout the observing run, did not impact our sensitivity.

The exposures were 300\,s long in both the on- and off-band filters. The observing strategy was to prioritize fields with the highest number of stars (as determined from the
number of 2MASS sources with $J<17$), and later fill in gaps.  In order to 
optimize the image subtraction, we repeated exposures until the full-width-at-half-maximum (FWHM) matched to within 0.1~pixel.  The seeing was typically good, with a median FWHM of 2.7 pixels (1\farcs2); our best exposures had values of 2.2 pixels (0\farcs9) and our worst, 4.1 pixels (1\farcs8).  Each exposure covered 29\farcm7 (N/S) $\times$ 29\farcm8 (E/W), at a scale of 0\farcs435~pixel$^{-1}$.  The 4096 $\times$ 4110 e2V CCD was read out through four amplifiers, resulting in
four raw images for each exposure.

As mentioned above, the actual count rates were as expected; our photometric zero-point, corresponding to a count rate of 1 $e^-$ s$^{-1}$, was equivalent to an AB magnitude of 19.55 at a typical airmass of 1.5.  One of our concerns before obtaining our first exposure was how much fringing to expect, as we were observing in the red through very narrow interference filters, with the possibility of OH lines present.  
%The potential had been likened to the possibility of an eccentric relative who might, or might not, show up at a dinner party.  
We were relieved to find no sign of fringing on our frames. The CCD did show much less spatial uniformity in the flat-fields, with many small-scale variations, but these were reproducible and flat-fielded out at much better than 1\%.  The sky values had a median value of 78 $e^-$ for the on-band filter, and 58 $e^-$ for the off-band, consistent with our expectation that the on-band would contain some OH contamination.

The reduction and calibrations procedures followed those of \citet{MasseyMCWRII}.  Ten bias frames were obtained each day.   Multiple twilight sky flats were taken through each filter, typically each night. Because the filters are so narrow, these exposures had to be started immediately at sunset.  The telescope was moved by 20\arcsec\ or more between each twilight exposure. The Swope CCD camera uses an iris type shutter, and as such a correction is needed for the short ($<$10~s) exposures of the twilight flats. (Details of this correction can be found in the lengthy footnote 4 to \citealt{MasseyMCWRII}.)  Several spectrophotometric standards were observed in order to set an approximate zero point for the photometry.

The basic reductions were done with {\sc iraf}, using modified versions of the reduction scripts written for the earlier Population I WR survey.  For each exposure, the processing steps included first determining the average bias overscan value for each of the four raw frames, and trimming off extraneous columns.  The frames read out through amplifiers 2--4 were next rotated and/or flipped to match the orientation of the first. A small linearity correction was applied to each image, depending upon which amplifier was used.  The four images were then merged into a single image, and
transposed to the conventional astronomical orientation (east to the left and north up).  The same process was applied to the bias frames and the sky flats.  The full bias frames were then averaged and subtracted from each of the other frames. 
The shutter correction mentioned above was then applied to each of the twilight flat-field exposures, and the flats combined filter-by-filter in such a way that any stars were filtered out (i.e., using {\sc iraf}'s {\it crreject} algorithm). The resulting flats were carefully examined to make sure that any stars present on the exposures had been removed, and then used to flat-field the science exposures.

The reduced images were then processed through an analysis pipeline, consisting of a series of {\sc iraf} scripts and {\sc fortran} programs. The end result was the addition of a world-coordinate system on each of the images utilizing our local installation of the astrometry.net package \citep{Lang10}.

In order to identify candidates that were significantly brighter in the on-band exposure than in the continuum, we compared the magnitude difference of each star to the combined expected photometric error associated with that difference.
The photometric error in each filter is computed automatically as part of the photometry, based upon the counts in the star, the counts in the sky, and the read-noise.  However, the signal-to-noise of the objects of interest is very high, and if we considered only photon noise and the (negligible) read-noise, most of our photometric errors would be in range of hundreds or even tens of milli-magnitudes.  Our experience has shown that taking these small errors at face value would result in many spurious candidates that were apparently statistically significant.  Indeed, there are other sources of uncertainties that affect the magnitude difference, such as large-scale flat-fielding issues; e.g., incompletely flattened dust donuts. We therefore adopted a minimum error in the magnitude difference of 0.02~mag, a somewhat conservative choice.  The ``significance level” of a detection was then evaluated by dividing the magnitude difference by the photometric error; the latter is
0.02~mag for most of the stars brighter than 16.0 in the continuum.   This significance level is simply a measure of how
believable the magnitude difference is between the two filters.  (Adopting this minimum error lowered the significance level of J0608 from the nominal 75$\sigma$ level mentioned earlier to 49$\sigma$.) We discuss the corresponding completeness limit below in Section~\ref{Sec-complete}.

\subsection{Identification of Candidates}

\subsubsection{Initial Candidate List}

With the reduced data in hand, we then created a list of candidates using both image subtraction and the photometric sigmas described above. Image subtraction was done using A. Becker's {\sc hotpants} software\footnote{http://web.ipac.caltech.edu/staff/fmasci/home/astro\_refs/HOTPANTSsw2011.pdf}, which allowed us to subtract the continuum (off-band) image from the on-band images. After this subtraction, the high significance candidates (those with higher flux in the on-band) were all that remained. An example is shown in Fig.~\ref{fig:postagestamps} for the field around J0608. 
However, the process is imperfect, and false positives do emerge. For example, bright stars, such as the one shown on the north edge of the Fig.~\ref{fig:postagestamps} images, sometimes leave a residual trace in the subtracted image. In this case, the photometric sigmas are particularly valuable, as they show the photometric variations between the on- and off-band images are generally close to zero for these bright stars, and thus they are not valid candidates. Another source of false positives in the photometry is the chance superposition of cosmic rays onto stars.
If this occurs in the on-band image, then the photometry will flag this as a potential candidate.  These cases were
readily eliminated by examining the image subtraction difference frames.  

Generally, confusion was not an issue
at these bright magnitudes, except in the most crowded regions, such as the R136 central cluster in 30 Dor. Therefore, the overall procedure for selecting the initial candidate list involved visually identifying the most promising candidates on the subtracted images and then cross-matching them with the photometric significance values. At the end of this process we were left with 53 candidates, all with photometric significance levels $>5 \sigma$.  This group included J0608.

\subsubsection{Astronomical False Positives}

Despite our best efforts, there were several types of astronomical false positives in the survey,
in the sense that the detected object is genuine, but probably lacks C\,{\sc ii} emission.  One subtle issue that
we failed to initially appreciate was that a few previously-cataloged broad-lined Population I WR stars appear in the survey as potential candidates.  The most numerous of these were WC4 WRs, the hottest and broadest line of the WC-subclass.  The red wing of an emission blend of C\,{\sc iv} $\lambda 7206$ and C\,{\sc iii} $\lambda 7212$ spills into our C\,{\sc ii} bandpass.  There is also a  C\,{\sc iv} $\lambda 7380$ line that contaminates
the continuum filter, but is significantly weaker, so an on-band excess remains.  Several previously-known WN4b stars were also recovered; these are particularly broad-lined, high-excitation members of the WN (nitrogen) subclass of WR.  Examination of our spectra of such stars (taken for a different project) shows that indeed there are weak, broad lines of N\,{\sc iv} at 7204\,\AA\ and 7247\,\AA\ that are also in the C\,{\sc ii} bandpass.   Our knowledge of the WR population of the LMC is believed to be complete thanks to the recent survey of \cite{NeugentMCIV}, and so such false positives are easily identified and removed from our list. 

Background active galactic nuclei at selected redshifts will also trigger false positives as their strong emission lines enter our on-band filter. We recovered the object 6dFGS~gJ042936.9-692653, a known emission-line galaxy with redshift 0.101 \citep{Jones09}, which shifts H$\alpha$ and [N\,{\sc ii}] emission (visible in the published spectrum) into our C\,{\sc ii} bandpass.

\subsubsection{Literature Vetting}
 
 After removal of the false positives noted above, our candidate list could clearly still contain previously-studied objects, as the LMC has been exhaustively cataloged photometrically, and to some extent spectroscopically. We performed a literature search to identify previously noted objects in the sample, and quickly identified three of our candidates with known objects. The well-studied R~CrB star HV\,2671, known for decades to have intense C\,{\sc ii} emission, was easily recovered. Also detected is J055825.96-694425.8, which has a spectrum with C\,{\sc ii} emission \citep{vanaarle}, and has been termed ``likely a hot proto-PN or PN” by \citet{hrivnak}. We do not yet have our own spectrum of this object, but it seems likely that it will also prove to be a late-type [WC] central star. Finally, the central star of the well-known LMC planetary nebula SMP~58 is weakly but still significantly detected in our photometry as a likely C\,{\sc ii} emitter; it is one of the three objects with new spectra that we
discuss below in \S3.2, where we refer to it as object 152-1.

Multiple photometric studies of the LMC have high-cadence, extended duration observations to enable searches for variable stars or lensing events. Curiously, several of these objects which are classified in the literature as either eclipsing binaries or Long Period Variables (LPVs) also appear as emission candidates in our survey, for reasons which are quite unclear. It is possible that in the case of the eclipsing binaries, photometric variability during the short interval between the centers of our on- and off-band exposures could explain the apparent on-band excess in some of these stars. For the LPVs, we speculate that sharp molecular bandheads may enhance or depress our on- or off-band filters, respectively, but a definitive explanation for both classes of variables must await spectroscopy.

All other candidates in the survey are present (generally without comment) in astrometric and/or photometric catalogs of the LMC. Therefore, none of these objects would appear to be a transient, previously undetected star. 

We plot in Fig.~\ref{fig:survey} the position of our 38 remaining candidates not otherwise explained above, along with the location of J0608. It is interesting to note that the observed distribution of these stars does not show a preference for the star-forming regions of the LMC. Therefore it seems likely that many or most of these objects are of an old population. However, further clarification of the nature of these candidates will obviously require spectroscopy.

\subsection{Completeness}
\label{Sec-complete}

Completeness in surveys such as ours is limited by two parameters: first, the emission-line fluxes, and second, the depth to which the photometry extends.   The magnitude difference between the on-band exposure and the off-band exposure will be basically a measure of the equivalent width of the emission (i.e., the line flux divided by the continuum flux) and not  the line flux {\it per se}.  A bright star will show a smaller magnitude difference than a faint star with the same amount of emission (as measured by the line fluxes).   However, the bright star will also have smaller photometric errors than the faint star, and thus by using a sigma cut-off we are in essence adopting a line-flux cut-off.   

In Fig.~\ref{fig:complete} we show the magnitude differences plotted as a function of AB magnitude in the continuum (off-band) filter for all of our candidates.  The position of J0608 is shown in green, and the three other stars discussed below, in red. We include the theoretical 5$\sigma$ line based upon the average sky values, the read-noise, and photon statistics as a function of magnitude.  Because we adopted 0.02~mag as the minimum error in the magnitude difference between the two filters, the curve is flat at brighter magnitudes.   Based upon our spectrum of J0608, the emission-line flux limit at 5$\sigma$ should correspond roughly to 2$\times10^{-15}$ ergs cm$^{-2}$ s$^{-1}$ at 16th mag. 

Our photometry extended down to objects whose peak brightnesses were about 10$\times$ greater than the noise in the background.  Examining the histograms of the resulting photometry shows that the faintest objects had a continuum AB magnitude of about 19; we were complete to a continuum AB magnitude of 17.5--18.0.  Note however that such a faint object would have to have a large magnitude difference (related to the equivalent width) of 0.3~mag to be considered a candidate.   We also note that there could be bright stars with the same amount of C\,{\sc ii} line fluxes as that of fainter stars that we ignored because of our insistence of a floor to the photometric error.  Whether or not this is a significant problem will be revealed only once we have spectroscopy of the many candidates that are hugging the 5$\sigma$ line at bright magnitudes.  

A salient question is whether our survey has the sensitivity to recover stars like J0608, and hopefully yet fainter objects with even weaker emission. The discussion above implies that this survey is designed to be sensitive to a line flux of C\,{\sc ii} emission that is $10\times$ weaker
than that in J0608, and down to a continuum magnitude that is $5\times$ fainter (i.e., $m_{\lambda7400}\sim18$). The preliminary spectrophotometric observations discussed below have thus far validated this conclusion.

\section{Spectroscopy}

\subsection{Observations}

During an unrelated observing program, on UT 2020 January 15 and 16 we obtained spectra of three high significance C\,{\sc ii} emission candidates, using the MagE spectrograph  \citep{Marshall08} on the 6.5\,m Baade Magellan telescope of the LCO.
MagE is an ideal instrument for this project, because it provides
complete wavelength coverage from 3150\,\AA, near the atmospheric cutoff, up to $\sim$10,000 \AA\,, at intermediate resolution (resolving power 
$\lambda$/$\Delta\lambda\sim$ 4100) in one single exposure, with minimal overhead.

We used a 1\arcsec\ slit oriented along the parallactic angle. Typical exposure times on the candidates were 600--1200\,s.
Prior to twilight we obtained the usual sets of bias frames, plus external 
quartz lamps for fringing correction in the red wavelength region. 
We also obtained the various recommended
sets of Xe-flash exposures, although those are not needed for our
reduction procedures.  Reductions were done with a
combination of {\sc iraf} {\it echelle} tasks and the {\it mtools} package originally
developed by J. Baldwin for the reduction of Magellan Inamori Kyocera Echelle (MIKE) data, and available online\footnote{http://www.lco.cl/telescopes-information/magellan/instruments/mike/iraf-tools/iraf-mtools-package}. 
This is the same procedure as applied to
previous MagE data (see, e.g., \citealt{Massey12} for a more detailed
description). 
Flux calibration was performed using a set of three or more spectrophotometric standard stars obtained during the same observing nights, and observed with the same instrument setup. Strong telluric features were removed using a spectrum of GD50 obtained during the same observing nights, by means of the {\sc iraf} {\it telluric} task. However, the process is not perfect, and some residual absorption, for example around the strong telluric A-band, remains.

The details of the three stars observed spectroscopically are given in Table~\ref{tab:phot}, and their locations in the LMC are marked on Fig.~\ref{fig:survey}. For simplicity, we have retained the candidate list names based upon
our survey field numbering scheme.  Note these are not the same fields as used in the LMC Population~I WR survey
described by \cite{MasseyMCWRI} and \cite{NeugentMCIV}. Table~\ref{tab:phot} also contains the corresponding data for J0608, for ease of comparison with the newly discussed stars. The photometry in the table should be regarded with caution, as many objects with C\,{\sc ii} emission are known to vary erratically (e.g., \citealt{miszalski}, M20). However, it is clear that all of these stars are fainter by 2--3~mag than the well-known (high-mass) WC stars of the LMC \citep{NeugentMCIV}. All the objects exhibit UV-excesses, and are also clearly (but anonymously) visible in Swift NUV UVOT images\footnote{http://mast.stsci.edu}.

\subsection{Comments on Individual Objects}

It is beyond the scope of this initial survey paper to analyze in individual detail the spectroscopically observed stars, although each has interesting and sometimes unique features. Rather, we give a brief description of the data, to preview the types of objects we expect to encounter when spectroscopy of the entirety of candidates is complete. The one-dimensional, reduced, telluric-corrected spectra described in this section are available electronically in FITS format.

\subsubsection{Object 153-1}

We find no previous specific mention of this star in the literature, other than a presence in multiple astrometric and photometric catalogs. The spectrum of object 153-1 is presented in Fig.~\ref{fig:153}. It is immediately clear that this star bears a striking resemblance to J0608 (Fig.~\ref{fig:filters}), and thus is precisely what our survey is designed to discover. The spectrum is dominated by intense C\,{\sc ii} emission, as well as dozens of weaker C\,{\sc ii} and numerous He\,{\sc i} emission lines, and common nebular forbidden transitions, although [O\,{\sc iii}] is missing, as is also true in J0608. The first few lower Balmer lines (H$\alpha$, H$\beta$, H$\gamma$) are also strongly in emission; presumably these are nebular in origin. Narrow absorption lines become common in the blue, many of them P~Cygni profiles on the C and He lines.
The lengthy line identification table for J0608 (M20, Table~1) may be used to readily identify almost all of the observed lines in this new object, although the Balmer emission and the important C\,{\sc iii} $\lambda5696$ line are slightly stronger in this object than in J0608. We also classify 153-1 as [WC11], although we do note the slight variation in different workers' classification criteria at this coolest end of the [WC] sequence (M20).  

The published {\it UBV} colors of this star are quite similar to those of J0608 (Table~\ref{tab:phot}), including a prominent UV excess visible in both the photometry and our spectrophotometry.  However, the near- and mid-infrared colors of 153-1 and J0608 differ substantially, but, presuming that the ejected dust from these systems is not necessarily spherically symmetric, an explanation of these differences may be as simple as variations in geometry and viewing angle.

There seems little doubt that this star is a member of the LMC. The observed heliocentric radial velocity in our spectrum, obtained from 21 emission lines excluding He\,{\sc i}, is $\sim$277 $\pm$ 8~km~s$^{-1}$, and agrees well with the mean for the LMC \citep{vandermarel}. There is limited evidence that the He\,{\sc i} emission velocities, 304 $\pm$ 15~km~s$^{-1}$, measured from 9 lines, may be systematically higher. The tabulated {\it Gaia} DR2 proper motion and parallax upper limit (Table~\ref{tab:phot}, \citealt{gaia}) are also consistent with {\it Gaia} data for LMC members (e.g., \citealt{helmi}; \citealt{vasiliev}).
As noted above, J0608 is located $\sim$4$^{\circ}$ from the center of the galaxy, which initially caused us to wonder if objects in this odd class avoid the densest regions of the LMC for some reason. The location of 153-1, much closer to the LMC center, removes this speculation.  

Through the kind assistance of N.~Hambly, we have examined a digitized version of the AAO/UKST SuperCOSMOS
H$\alpha$ Survey film \citep{parker} which includes 153-1. No nebulosity is apparent, and if it exists, it must be $\lesssim$2\arcsec~in extent. The $\sim$15$\arcsec$ extent of the PN surrounding the Galactic prototype [WC11] star  CPD $-56^\circ$8032 \citep{chesneau} implies that a similar nebula at the LMC distance would be inconspicuous in ground-based images. There is also no imaging evidence of a planetary nebula around J0608 (M20), but one is likely present in both cases, due to the appearance of prominent forbidden lines of multiple species, and also the known association of Galactic [WC11] stars with PNe.  However, while J0608 exhibits JHK colors compatible with the compendium of PNe with late-[WC] central stars given by \citet{Muth20} and the selection criteria of \citet{akras}, 153-1 does not. The WISE mid-infrared colors of both of these stars do lie roughly in the ranges for this spectral type listed by \citet{Muth20}, although the general scatter of WISE colors for these objects is quite large. The catalog of \citet{vanaarle} flags this star as a post-AGB candidate, based on {\it Spitzer} photometry.

Finally, we note that this object should not be confused with 2MASS J05312189-7040454, which curiously has the virtually identical right ascension, but is $23'$ further south, and is a well-known PN, SMP~73 \citep{SMP}.

%\eject

\subsubsection{Object 152-1}

This is the central star of a previously-known, well-studied planetary nebula SMP~58 \citep{SMP}. The nebula has been the subject of dozens of studies over many decades, but there are only limited previous comments in the literature on the central star, which has been classified as [WC4] by \citet{monk}. The presence of weak C\,{\sc ii} $\lambda4267$ (but not the $\lambda\lambda7231, 7236$ doublet) emission has also been tabulated (e.g., \citealt{leisy}). 

Our spectrum of object 152-1, presented in Fig.~\ref{fig:152}, is more complex than previous descriptions. It is dominated by several dozen very strong, common emission lines of PNe. The strongest nebular lines are badly saturated in the 1200\,s exposure needed to reach the stellar continuum, so we also obtained additional spectra with integrations of 300\,s and 50\,s, and insert these flux-calibrated data from this shortest exposure for the strongest emission lines in the figure. In order to more clearly simultaneously display data for both the nebula and the central star, this plot is semi-logarithmic.

Our spectrum displays C\,{\sc ii} $\lambda\lambda7231, 7236$ emission at the LMC velocity, as expected from our selection process, as well as numerous other C\,{\sc ii} emission lines. In a spectroscopic study of the nebula, \citet{stan02} comment ``an emission feature at about 6577\,\AA\, is probably stellar;" this is surely the C\,{\sc ii} $\lambda\lambda6578, 6582$ doublet, which we strongly detect. Although our spectrum contains contributions from both the PN and its central star, we believe the C\,{\sc ii} in this case is stellar, as is the case in the late [WC] stars. Although \citet{peimbert} note that the C\,{\sc ii} $\lambda 4267$ is sometimes seen in emission nebulae due to recombination, the
multiple, strong C\,{\sc ii} lines in our data are not typical of nebular spectra.  Indeed, M20 point out that for J0608 (and by inference for other late [WC] stars), the observed C\,{\sc ii} lines are stronger, sometimes greatly so, than permitted by simple recombination theory, 
requiring instead formation in the stellar wind of the central star itself.

Numerous strong, narrow He I emission lines are also present,
some with prominent  P~Cyg profiles, again suggesting a stellar, rather than nebular, origin. The H$\delta$ emission also has a blue absorption wing. Our measured heliocentric radial velocity from 43 emission lines, 278 $\pm$ 7~km~s$^{-1}$, agrees with the 295 $\pm$ 23~km~s$^{-1}$ nebular velocity of \citet{reid06} and confirms LMC membership.

The narrow C\,{\sc ii} and He\,{\sc i} emission, combined with minimal (or no) C\,{\sc iii} $\lambda5696$ emission, would on the surface
suggest a late [WC] type for the central star, likely [WC11], quite opposite to the [WC4] classification in the literature.  However, in agreement with \citet{monk}, broad  C\,{\sc iv} $\lambda4650$, He\,{\sc ii} $\lambda468$6, and C\,{\sc iv} $\lambda\lambda5801, 5812$ emission
(all at the LMC velocity), along with weak O\,{\sc v} $\lambda5592$ emission, are also prominent in our spectrum, instead consistent with a very early [WC] type,
i.e., [WC4] as shown for the central star of NGC~6751 by \citet{chu}, and in other examples
presented by \citet{crowther}. With the lack of C\,{\sc iii}, the presence of both strong C\,{\sc iv} and C\,{\sc ii} requires
two distinct emission regions.  This is also consistent with the difference in line widths, as the hot component (He\,{\sc ii}, C\,{\sc iv}, O\,{\sc v}) have very large line widths (the observed FWHM of 
C\,{\sc iv} $\lambda$5806 is 23\,\AA\, or 1200~km s$^{-1}$), while the cooler components (He\,{\sc i},  C\,{\sc ii}) have line widths which are unresolved at our spectral resolution; the C\,{\sc ii} $\lambda7231$ component
has a measured FWHM of 1.8\,\AA, the same as our instrumental resolution at that wavelength.   Could this be a [WC4]+[WC11]
binary?  If so, this would be the first such object known.  It would require the two
stars to be of comparable visual brightness, but little is known about the luminosities of such objects \citep{demarco02}. Binarity has been suggested for a few late-type [WC] stars (\citealt{hajduk}, \citealt{miszalski}). Of course, a two [WC] star system would require both components reaching this relatively short-lived stage simultaneously, either through single- or binary-evolutionary channels.

Alternative explanations for the cooler component, such as a nebular or circumstellar origin, would have to account for the very numerous permitted C\,{\sc ii} emission lines, not commonly seen in nebular spectra, and the P~Cyg components of He\,{\sc i}. Follow-up radial velocity monitoring is planned for the next observing season.

\subsubsection{Object 233-1}

The star 233-1 appears in numerous astrometric and photometric catalogs; there is an intense UV excess, including a detection in the NUV by the XMM Optical Monitor telescope \citep{page}. We are not aware of any previous mention of the spectrum, which we present in Fig.~\ref{fig:233}. It is immediately clear that this object is entirely different than the two [WC] stars discussed above. The C\,{\sc ii} doublet which led to our selection of this star is evident in emission at the LMC velocity, and is fully resolved in our data, leaving no doubt as to the identification. However, there are no other prominent emission lines, including the expected C\,{\sc ii} $\lambda$4267. Most striking, however, is the extraordinarily dense thicket of narrow absorption lines, particularly in the blue
part of the spectrum, unresolved at our $R\sim4100$ resolving power.   The colors are very blue, which would normally imply a high
effective temperature, and we would therefore expect both a sparse absorption line spectrum and relatively broad lines.

The numerous observed lines make identifications problematic, lacking any {\it a priori} knowledge of physical conditions and allowing for various velocities, especially when considering species of relatively common elements and ionization states such as Fe\,{\sc ii}, which have hundreds of lines in this region and thus will likely have many plausible but coincidental matches. To obtain a start on absorption line identifications that are simultaneously consistent for a large number of features, we constructed arrays of the rest wavelengths of the first 500 permitted lines between 4000 and 5000 \AA\, of of C\,{\sc ii}, C\,{\sc i}, Ne\,{\sc ii}, Ne\,{\sc i}, O\,{\sc ii}, O\,{\sc i}, He\,{\sc ii}, He\,{\sc i}, and Si\,{\sc iii}, using the atomic line list 
v2.04\footnote{http://www.pa.uky.edu/$\sim$peter/atomic/}. Each of these arrays was shifted in velocity over the $-500$ to +500~km~s$^{-1}$ range in 2~km~s$^{-1}$ steps, and cross-correlated with the observed spectrum, searching for peaks in the correlation function. A very strong peak is found at $\sim$+280~km~s$^{-1}$, consistent with the LMC velocity, for O\,{\sc ii}, and slightly weaker but significant correlations at this velocity for He\,{\sc i}, He\,{\sc ii}, Si\,{\sc iii}, N\,{\sc ii}, and possibly C\,{\sc ii}. 
 
The SuperCOSMOS
H$\alpha$ Survey film again shows no nebulosity, also with a limit of  $\lesssim$2\arcsec~extent. We do note that this star lies just outside the eastern-most edge of the bright H$\alpha$ nebulosity associated with the extensive emission complex DEM~76 \citep{davies}.

This object appears to be a hot, H-poor post-AGB star of some sort, but with quite unique features, not exactly  matching other cases in the literature. We note that J0608 also shows a myriad of narrow absorption lines in the blue, including numerous O\,{\sc ii} features (Fig.~\ref{fig:filters}; M20). It is possible that 233-1 is caught in transition from a late [WC] star to a hot white dwarf. This speculation is supported by the numerous narrow lines in the blue of both stars, and also the odd appearance of the C II doublet as the only prominent emission feature in 233-1. The object is certainly worthy of further study.

\section{Summary and Conclusions}

A narrowband imaging survey of $\sim$50~deg$^2$ of the LMC, designed to detect C\,{\sc ii} emission line stars with $m_{\lambda7400}\leq18$, has identified 38 candidates. We present spectra of three of the candidates, thus verifying that the selection technique is effective. 
Although all of the objects have UV-excesses, the variety of observed visible and infrared colors of these stars, even among objects with quite similar spectra such as 153-1 and J0608 (Table~\ref{tab:phot}), implies it would be difficult to systematically discover these objects through color selection alone. Indeed, based on WISE colors, the [WC11] star J0608 is selected as an AGN by \citet{secrest} and \citet{paine}, in a list said by the former authors to be pure of stellar contamination at the $10^{-5}$ level.

\begin{figure}
\plotone{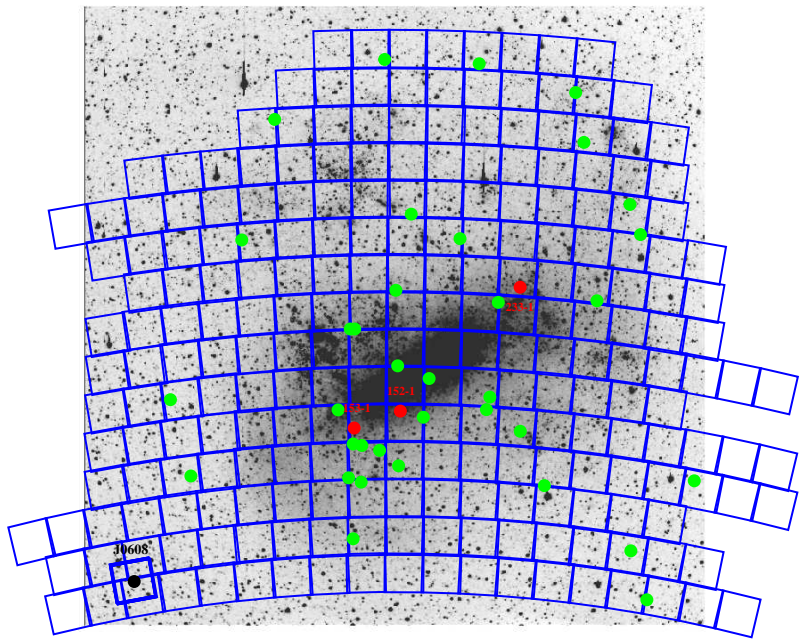}
\caption{\label{fig:survey} The LMC C\,{\sc ii} Survey fields.  ${\it Blue\, squares}$: the 246 fields included in the survey.  Each field is 29\farcm7 (N/S) $\times$ 29\farcm8 (E/W) on a side.  ${\it Black~dot}$: the location of the LMC [WC11] star J0608; ${\it green~dots}$: stars with observed on-band photometric excesses; ${\it red~dots}$: candidates spectroscopically observed in the initial reconnaissance reported here. The underlying LMC image is the R-band ``parking lot" frame described by \citet{Bothun}. North is up and east to the left.}
\end{figure}

\begin{figure}
\plotone{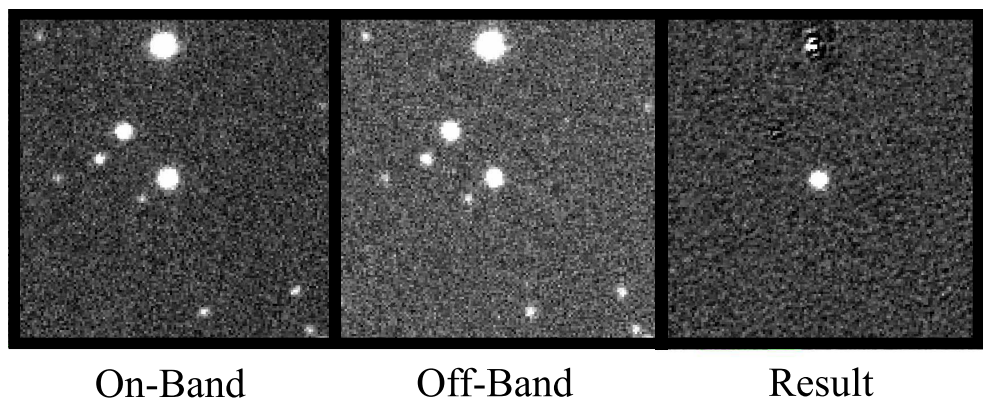}
\caption{\label{fig:postagestamps} An example of the image subtraction technique used to select candidates with C\,{\sc ii} excesses. The fields shown are $1'$ $\times$ $1'$ squares centered on the known strong C\,{\sc ii} emitter J0608; north is up and east to the left. After observing each field through an on-band and off-band image, the latter was then subtracted from the former, so that any object with a higher on-band flux is prominent. Note the marked excess for J0608 in the subtracted frame ({\it right}). However, the image subtraction technique is imperfect, especially for brighter stars, as can be seen by the residual left by the bright object at the north edge of the field.}
\end{figure}

\begin{figure}
\plotone{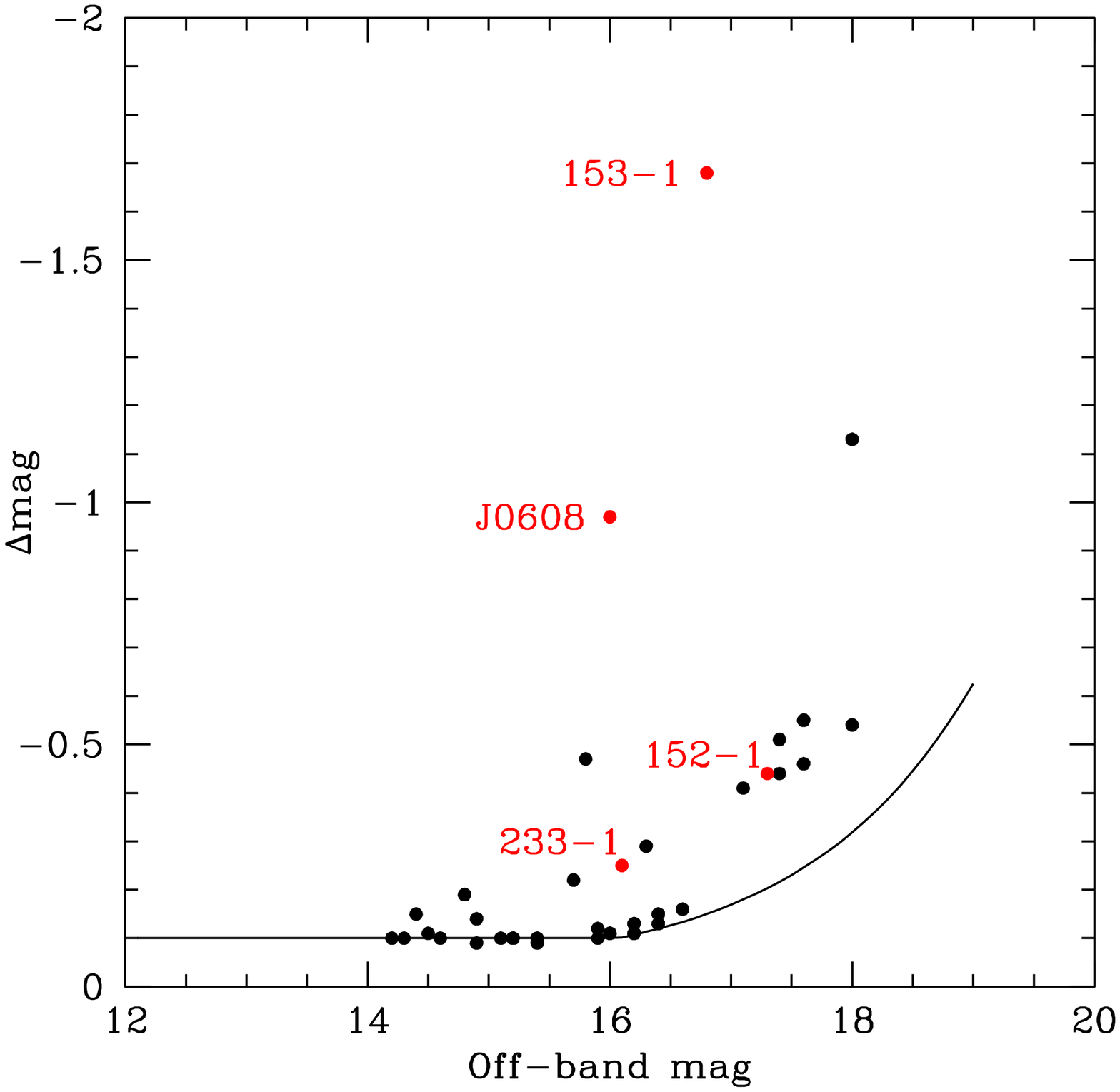}
\caption{\label{fig:complete} Completeness of our survey.  The magnitude difference $\Delta$m (C\,{\sc ii} on-band magnitude minus continuum off-band magnitude) is plotted against the AB continuum magnitudes for our candidates. {\it Red points}: stars with spectra confirming the presence of C\,{\sc ii} emission}. The theoretical 5$\sigma$ limit is shown by the curve. All four of the confirmed C\,{\sc ii} emission line stars are found at high significance levels; further spectroscopy is needed to establish the nature of the other candidates, particularly those with lower significance values.

\end{figure}

\begin{figure}
\plotone{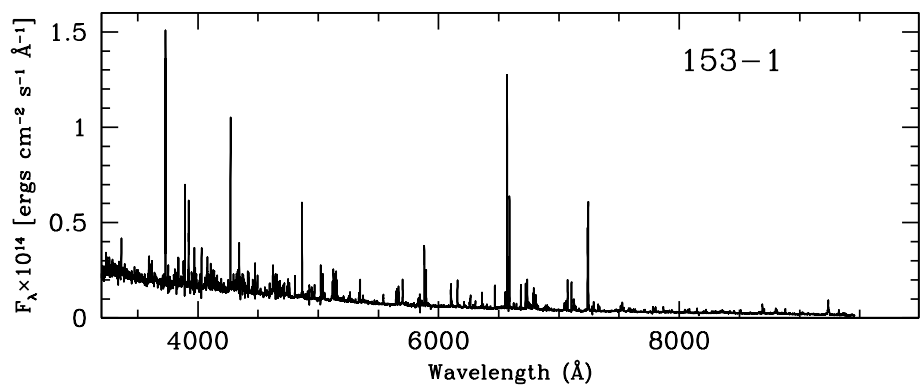}

\vskip -175pt

\caption{\label{fig:153} The spectrum of star 153-1. Note the strong resemblance to that of the [WC11] star J0608 (Fig.~\ref{fig:filters}, upper panel). In addition to Balmer emission and the common nebular forbidden lines, both spectra are dominated by numerous C\,{\sc ii} emission lines. The one-dimensional, reduced, telluric-corrected spectrum used in this figure is available electronically in FITS format.}

\end{figure}

\begin{figure}
\plotone{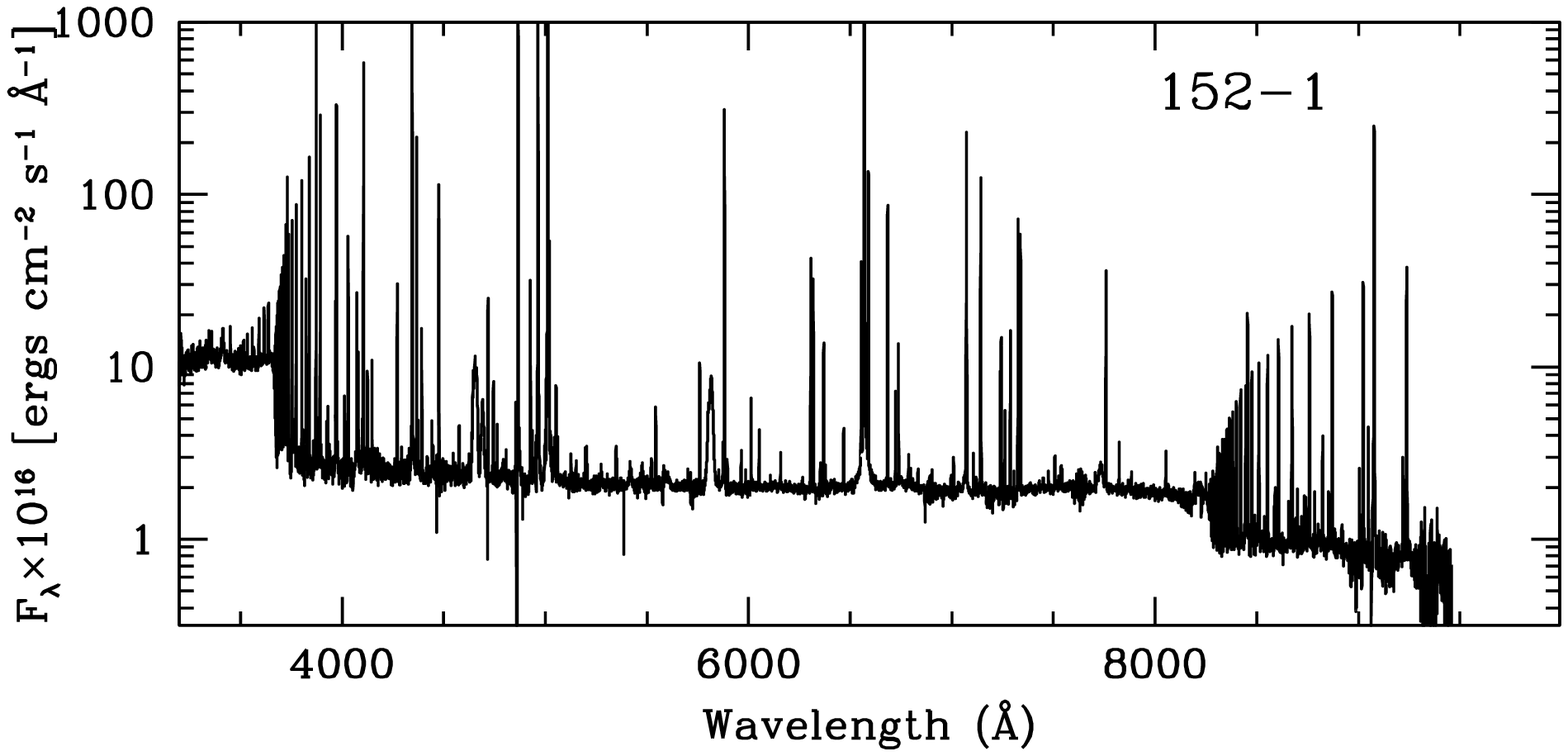}

\vskip -150pt

\epsscale{0.48}
\plotone{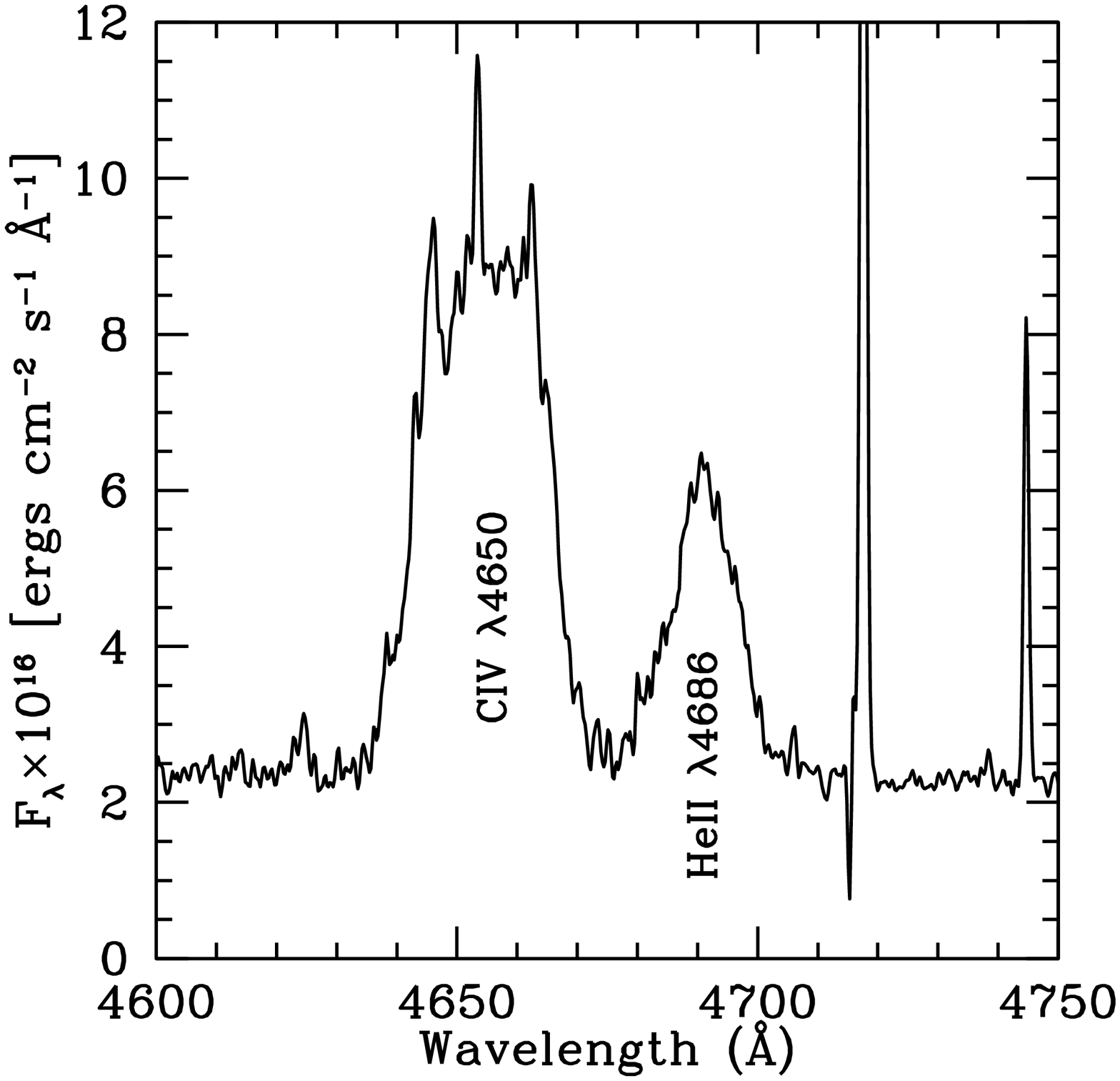}
\plotone{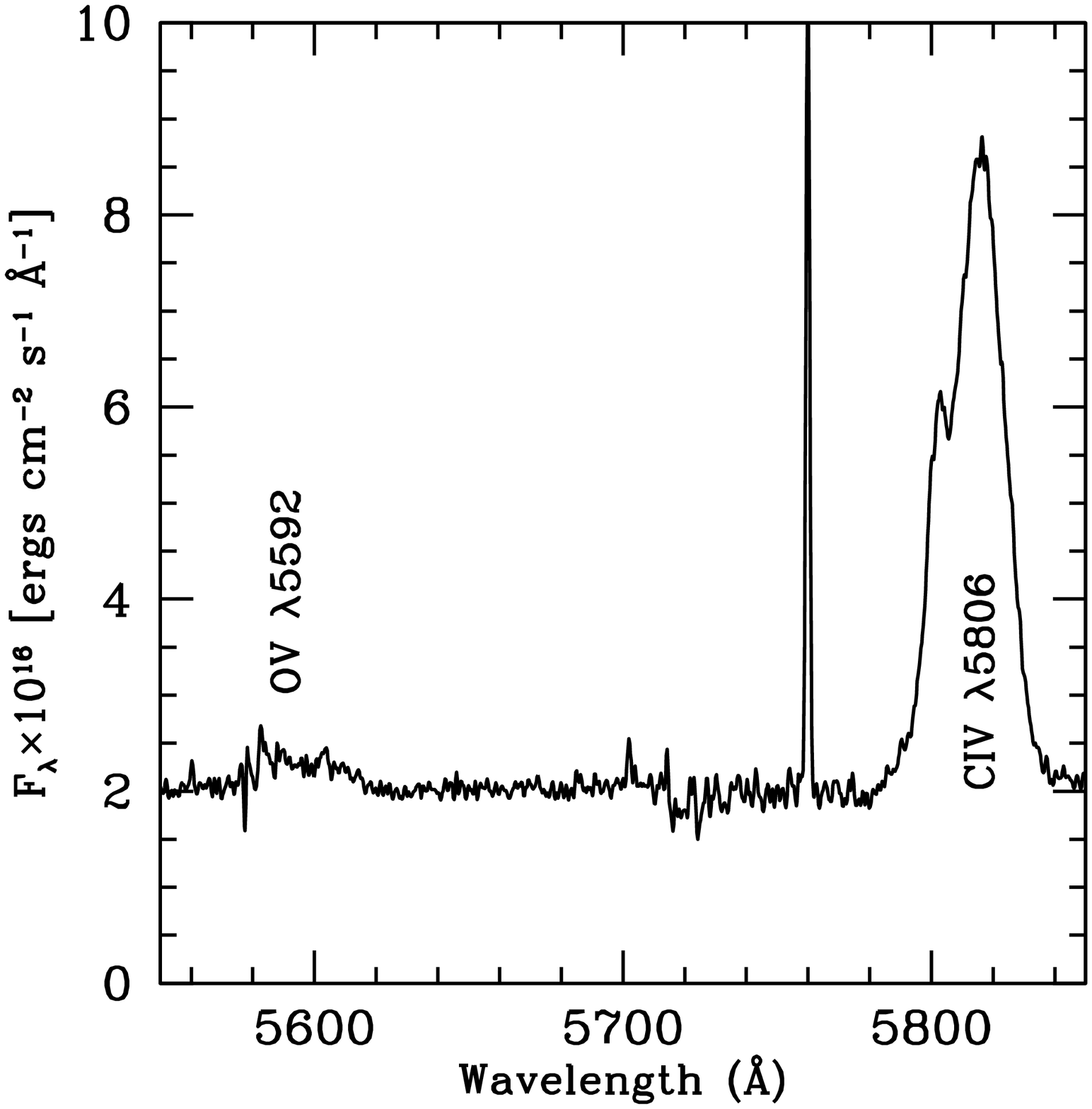}

\vskip 50pt

\caption{\label{fig:152} ${\it Upper}$: The spectrum of star 152-1. The flux axis is logarithmic, chosen to enable simultaneous display of the multiple, very strong emission features of the known planetary nebula SMP~58 as well as those of the central star. The confluence of both the Balmer and Paschen series is also prominent. ${\it Lower}$: Enlargements of two unique parts of the spectrum. Although many of the narrow C\,{\sc ii} and He\,{\sc i} emission lines of the [WC11] stars are present, broad emission of C\,{\sc iv} $\lambda$4650, He\,{\sc ii} $\lambda$4686, O\,{\sc v} $\lambda 5592$, and C\,{\sc iv} $\lambda\lambda$5801, 5812 is also evident, indicative of a much hotter star. Note the prominent P~Cyg profile of the He\,{\sc i} $\lambda$4713 line in the lower left panel, which argues for a stellar rather than nebular origin. The one-dimensional, reduced, telluric-corrected spectrum used in this figure is available electronically in FITS format.}

\end{figure}

\begin{figure}
\plotone{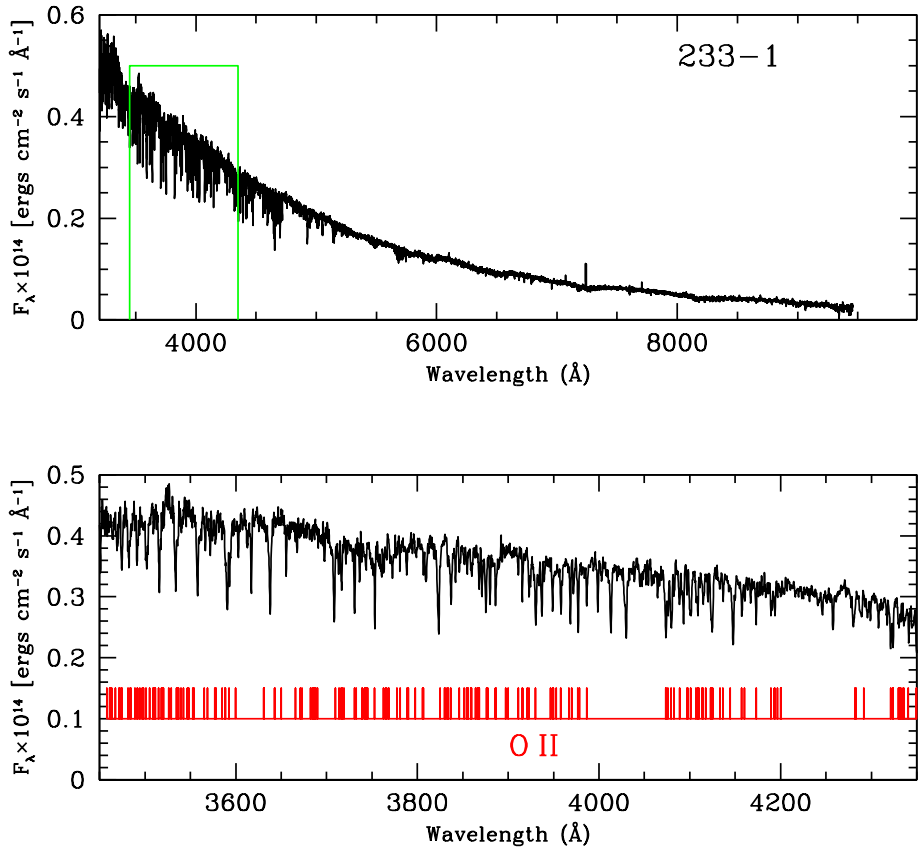}

\caption{\label{fig:233} {\it Upper}: The spectrum of star 233-1. The sole obvious emission is the C\,{\sc ii} $\lambda\lambda$7231, 7236 doublet that triggered selection in our survey. A marked UV-excess continuum is also evident. The green box denotes the area shown in the lower figure. {\it Lower}: An enlargement of a section of the blue portion of the spectrum, in order to highlight the extraordinarily dense set of absorption lines, unresolved at our $\sim$1~\AA\, instrumental resolution. The red bars are the O\,{\sc ii} transitions in this spectral region. There are a large number of matches, although the correspondence is imperfect. As discussed in the text, there are also likely contributions from He\,{\sc i}, He\,{\sc ii}, Si\,{\sc iii}, N\,{\sc ii}, and possibly C\,{\sc ii}. Note the absence of any appreciable C\,{\sc ii} $\lambda$4267 emission, which in the late [WC] stars is comparable or greater than the strength of the $\lambda\lambda$7231, 7236 doublet. The one-dimensional, reduced, telluric-corrected spectrum used in this figure is available electronically in FITS format.}

\end{figure}

More quantitative estimates of our completeness, and limits in both stellar flux and C\,{\sc ii} emission equivalent width, must await spectroscopic verification of the candidates of the lowest significance. However,
the observed number and the space distribution of the candidates (Fig.~\ref{fig:survey}) already permit several interesting inferences, even lacking spectral classifications for the majority of the stars. The objects are not concentrated in the star-forming regions of the galaxy, and so are probably comprised of one (or more) old populations. The odd location of the initially discovered object, J0608, in the far outskirts of the visible galaxy, proves not to be typical.

The LMC contains $\sim10^3$ planetary nebulae \citep{jacoby, reid}. Therefore, the relatively small number of C\,{\sc ii} emission candidates reported here, despite the wide area of the survey, implies that, even given the difficulty in their discovery, mid-to-late type [WC] stars likely do not dominate the population of planetary nebulae central stars. This conclusion, even if expected, would be difficult to defend lacking a specialized survey such as the one described here.

These odd stars are interesting for a variety of reasons, including estimation of stellar carbon abundances. M20 have pointed out that there are likely selective population mechanisms at work to explain the observed strength of the $\lambda\lambda7231/36$ doublet, so past abundance inferences which utilize the strong C\,{\sc ii} lines must be viewed with caution. Thus, given the modest current total number of mid-to-late [WC] stars, both Galactic and extragalactic, even a small addition to the known sample may be helpful in understanding the atomic physics involved. As the luminosities of the known Galactic late [WC] stars are quite uncertain, the accurate luminosities available for these LMC objects are also valuable. Our very preliminary spectroscopic reconnaissance has added two more stars to this list, one of which may be an unusual [WC] binary. We plan to followup the remainder of our candidate list spectroscopically, and anticipate adding further interesting objects.

\acknowledgements
We thank the Mt.\ Cuba Astronomical Foundation for their generous support, which enabled purchase of the interference filters. We are also grateful to the anonymous referee for useful comments which have improved the manuscript, to Rob Fesen for discussions about modern interference filters, and to Dick Stewart of Chroma Technology for his help with our unique specifications.  We thank Dick Joyce and George Jacoby for their assistance in characterizing the delivered filters. Useful correspondence with John Hillier, Thomas Kupfer, George Jacoby, and Howard Bond is also acknowledged. We are grateful to Nigel Hambly for supplying the SuperCOSMOS images. This work was supported by the National Science Foundation (NSF) under AST-1612874, and NSF IGERT grant DGE-1258485. This work has made use of data from the European Space Agency (ESA) mission
{\it Gaia} (\url{https://www.cosmos.esa.int/gaia}), processed by the {\it Gaia}
Data Processing and Analysis Consortium (DPAC,
\url{https://www.cosmos.esa.int/web/gaia/dpac/consortium}). Funding for the DPAC
has been provided by national institutions, in particular the institutions
participating in the {\it Gaia} Multilateral Agreement.  The observations reported here were possible thanks excellent support from the LCO technical and support staff. We thank Ian Thompson and Leopoldo Infante for enabling the Swope observing time. As is the case at most ground-based observatories, operations at LCO are currently suspended due to the COVID-19 crisis.  We look forward to the next time we can be together again.

\facilities{Magellan:Baade, Swope}

%\eject
	
\bibliographystyle{apj}
\bibliography{WC11.bib}

 \begin{deluxetable}{ccccc}[h]
 \tabletypesize{\small}
 \tablecaption{Spectroscopically Verified LMC C\,{\sc ii} Emission Stars \label{tab:phot}}
 
 \tablehead{
\colhead{Parameter}
&\colhead{153-1}
 &\colhead{152-1}
 &\colhead{233-1}
 &\colhead{J0608}\\
 }
 \startdata
$\alpha$ & 05 31 21.77  & 05 24 20.77 &  05 07 38.90  & 06 08 19.94  \\
$\delta$ & -70 17 40.0  & -70 05 01.4   & -68 26 06.1  & -71 57 37.4 \\
$\mu_{\alpha}$ (mas yr$^{-1}$) & $1.95 \pm 0.08$ & $1.78 \pm 0.16$ & $1.79 \pm  0.07$ & $1.88 \pm 0.06$ \\
$\mu_{\delta}$ (mas yr$^{-1}$) & $0.48 \pm 0.10$ & $-0.09 \pm 0.17$ & $-0.22 \pm 0.09$ & $1.03 \pm 0.07$ \\
$\pi$ (mas) & $-0.04 \pm 0.04$ & $-0.15 \pm 0.08$ & $0.09 \pm 0.04$ & $0.00 \pm 0.03$ \\
$m_{\lambda 7410}$  & 16.8  &  17.3 & 16.1 & 16.0 \\
$\Delta m$\tablenotemark{a} & -1.68 & -0.44 & -0.25 & -0.97 \\
$\sigma$\tablenotemark{b} & 62 & 10 &  12 & 49 \\
2MASS & 05312172-7017394  & 05242076-7005015   &  05073893-6826061 & 06081992-7157373  \\
$V$ & 16.27  & 15.37 & 15.65  & 15.64 \\
$(B-V)$ & -0.16  & 0.87 &  -0.03  & 0.02 \\
$(U-B)$ & -0.75  & -0.30   &  -1.09  & -0.76 \\
$J$ & 16.19 & 15.73 & 15.79 & 15.48 \\
$(J-H)$ & 1.41 & 0.17 & -0.15 & 0.13  \\
$(H-K)$ & -0.35 & 1.00 & 0.09 & 0.79 \\
$W2$ & 13.78 & 11.14 & 14.17 & 11.12 \\
$(W1-W2)$ & 0.47 & 1.42 & -0.20 & 2.03 \\
$(W2-W3)$ & 6.80 & 4.47 & 0.63 & 4.51 \\
$(W3-W4)$ & 2.94 & 2.56 & 3.70 & 2.67 \\
\enddata
 
\tablenotetext{a}{magnitude difference, on-band minus off-band filter}
\tablenotetext{b}{Significance level of detection ($\Delta m$ divided by photometric error)}
\tablecomments{Positions (J2000), proper motions ($\mu$), and parallax ($\pi$) are from {\it Gaia} DR2 \citep{gaia}, UBV photometry from \citet{zaritsky}, JHK photometry from 2MASS \citep{2MASS}, WISE mid-infrared data (``W") from \citet{wright}.}

\end{deluxetable}

%Figure 1

%Figure 2

%Figure 3

%Figure 4

%Figure 5

%Figure 6

%Figure 7

\end{document}